\documentclass[10pt]{iopart}

\usepackage{booktabs}
\usepackage{xy}
\usepackage{bbold}
\usepackage{amsmath}

\input xy
\xyoption{all}

\def\1{\mathbb 1}

\numberwithin{equation}{section}

\usepackage{hyperref}  % COMPILE MORE TIMES for HYPERTEXT REFERENCES

\begin{document}

\title{On the linear equation method for the subduction problem in symmetric groups}

\author{Vincenzo Chilla}

\address{Dipartimento di Fisica ``Enrico Fermi'', Universit\`a di Pisa and Sezione INFN  - Largo Bruno Pontecorvo 3, 56127 Pisa, Italy}
\ead{chilla@df.unipi.it}

\begin{abstract}
We focus on the tranformation matrices between the standard Young-Yamanouchi basis of an irreducible representation for the symmetric group $S_n$ and the split basis adapted to the direct product subgroups $S_{n_1} \times S_{n-n_1} $. We introduce the concept of subduction graph and we show that it conveniently describes the combinatorial structure of the equation system arisen from the linear equation method. Thus we can outline an improved algorithm to solve the subduction problem in symmetric groups by a graph searching process. We conclude observing that the general matrix form for multiplicity separations, resulting from orthonormalization, can be expressed in terms of Sylvester matrices relative to a suitable inner product in the multiplicity space. \\
\phantom{a}\\
PACS numbers: 02.10.Ox, 02.10.Ud, 02.20.Hj.\\
Mathematics Subject Classification: 05E10, 15A06, 20C30.  
\end{abstract}

%=======================================================================================================

%\tableofcontents

%*******************************************************************************************************

\section{Introduction}
Subduction coefficients for symmetric groups were first introduced in 1953 by Elliot {\it et al}~\cite{Elliot} to describe the states of a physical system with $n$ identical particles as composed of two subsystems with $n_1$ and $n_2$ particles respectively ($n_1 + n_2 = n$). Later these coefficients assumed a central role in the so-called {\it Wigner-Racah calculus} via Schur-Weyl duality~\cite{Kramer, Vanagas, Haase}. In fact, the subduction coefficients are directly related to the coupling ($3j$) and recoupling ($6j$) coefficients of unitary groups which are often useful for simplifying many-body calculations in quantum or nuclear physics and chemistry. In particular, the $6j$ of the unitary groups can be expressed as sum of products of such coefficients~\cite{Chenbook, Pan1}.
\par
Since Elliot {\it et al} (1953), many techniques have been proposed for calculating the subduction coefficients, but the investigation is until now incomplete. The main goal to give explicit and general closed algebraic formulas has not been achieved. Only some special cases have been solved~\cite{Kaplan2, Rao, McAven}. There are numerical methods~\cite{Horie, Kaplan1, Chen1} which are used to approach the issue, but no insight into the structure of the trasformation coefficients can be obtained. Another key outstanding problem is to resolve multiplicity separations~\cite{Butler} in a systematic manner, indicating a consistent choice of the indipendent phases and free factors. In~\cite{McAven1, McAven2} a breakthrough was made about this; however, the authors abandon the aim to obtain an algebraic solution and prefer a combinatorial recipe.  
\par
In this paper we come back to an algebraic approach to the subduction problem in symmetric groups and we analyze in detail the {\it linear equation method}~\cite{Chen}, an efficient tool for deriving algebraic solutions for fixed values of $n_1$ and $n_2$. In section $2$ we provide some background and describe the method, giving the general structure of the resulting equation system ({\it subduction matrix}). In section $3$ we introduce the {\it subduction graph} and in section $4$ we relate it to the subduction matrix. The graph provides a graphic description of a minimal set of equations which are sufficient to obtain the trasformation coefficients. We find the solution space as an  intersection of suitable linear subspaces of $\mathbb{R}^{N} \otimes \mathbb{R}^{N_1 N_2}$, where $N$, $N_1$ and $N_2$ are the dimensions of the irreducible representations involved in the subduction. In section $5$ we give the general orthonormalized form for the coefficients and we discuss the choice of phases and free factors governing the multiplicity separations. We summarize our results in section $6$.

%*******************************************************************************************************

\section{The linear equation method: background}
\subsection{Standard and split basis}
The irreducible representations (irreps) of the symmetric group $S_n$ may be labelled by partitions $[\lambda]$ of $n$, i.e. sequences $[\lambda_1, \lambda_2, \ldots, \lambda_h]$ of positive integers such that $\sum_{i=1}^h \lambda_i = n$ and the $\lambda_i$ are weakly decreasing. A partition $[\lambda]$ is usually represented by a Ferrers diagram (or Young diagram) obtained from a left-justified array with $\lambda_j$ boxes on the $j$th row and with the $k$th row below the $(k-1)$th row. Standard Young tableaux are generated by filling the Ferrers diagram with the numbers $1, \ldots, n$ in such a way that each number appears exactly once and the numbers are strictly increasing along the rows and down the columns. An orthonormal basis vector of an irrep associated to the partition $[\lambda]$ may be labelled by a standard Young tableau. Such a basis corresponds to the Gelfand-Tzetlin chain $ S_1 \subset S_2 \subset \ldots \subset S_n$ and is usually called the \emph{standard basis} of $[\lambda]$. We denote this basis by \emph{$S_n$-basis}~\cite{McAven}.
\par
An alternative orthonormal basis for $[\lambda]$ is the \emph{split basis}, denoted by \emph{$S_n-S_{n_1,n_2}$-basis}~\cite{McAven}, with $n_1 + n_2 = n$. By definition, such a basis breaks $[\lambda]$ (which is, in general, a reducible representation of the direct product subgroup $S_{n_1} \times S_{n_2}$) in a block-diagonal form: 
\begin{equation}
[\lambda]=\bigoplus_{[\lambda_1],[\lambda_2]} \{\lambda; \lambda_1, \lambda_2\} \ [\lambda_1]\otimes[\lambda_2],
\end{equation}
where [$\lambda_1]$ and $[\lambda_2]$ are irreps of $S_{n_1}$ and $S_{n_2}$ respectively, and $\{\lambda; \lambda_1, \lambda_2\}$, the Clebsch-Gordan series, counts the number of times (\emph{multiplicity}) that the irrep $[\lambda_1] \otimes [\lambda_2]$ of  $S_{n_1} \times S_{n_2}$ appears in the decomposition of $[\lambda]$. 
\par   
The irreps of the subgroup $S_{n_1} \times S_{n_2}$ may be labelled by pairs $(\alpha, \beta)$ of Ferrers diagrams, with $\alpha$ corresponding to an irrep of $S_{n_1}$ and $\beta$ to an irrep of $S_{n_2}$. In the same way, each element of the basis is labelled by pairs of standard Young tableaux.  

%-------------------------------------------------------------------------------------------------------

\subsection{Symmetric group action on standard and split basis} 
The symmetric group $S_n$ of $n$ elements is generated by the $n-1$ transpositions $g_i$ each one interchanges the elements $i$ and $i+1$.
\par 
Given a standard Young tableau $m$, we define the action $g_i(m)$ in the following way: if the tableau obtained from $m$ interchanging the box with $i$ and the box with $i+1$ (keeping the other elements fixed) is another standard Young tableau $m^{(i)}$, we set $g_i(m)=m^{(i)}$; else $g_i(m)=m$.
\par
The $g_i$ acts on the standard basis vectors $| \lambda, m \rangle$ of the irrep $[\lambda]$ as follows~\cite{Chen}:
\begin{equation}
g_i | \lambda, m \rangle =
\left\{
\begin{array}{cc}
 \frac{1}{d_i(m)} | \lambda, m \rangle + \sqrt{1-\frac{1}{{d_i(m)}^2}} \ | \lambda, g_i(m) \rangle & \text{if $g_i(m) \ne m$}  \\
 | \lambda, m \rangle &  \text{if $g_i(m) = m$}
\end{array}
\right. ,
\label{actstd}
\end{equation}  
where $d_i(m)$ is the usual \emph{axial distance} from $i$ to $i+1$ in the standard Young tableau $m$~\cite{Fulton} .
\par 
The explicit action of the generators $g_i$ ($i \neq n_1$ because $g_{n_1}$ is not a generator of $S_{n_1} \times S_{n_2}$) on the elements of the $S_n-S_{n_1,n_2}$-basis directly follows from (\ref{actstd}). In fact we have
\begin{equation}
g_i | \lambda_1, \lambda_2 ; m_1, m_2 \rangle=
\left\{
\begin{array}{cc}
 (g_i |\lambda_1, m_1 \rangle) \otimes |\lambda_2, m_2 \rangle & \text{if $1 \le i \le n_{1}-1$ } \\
|\lambda_1, m_1 \rangle \otimes (g_i|\lambda_2, m_2 \rangle) & \text{if $n_1+1 \le i \le n-1$}
\end{array}
\right.
\label{actsplit}.
\end{equation} 
Then, from (\ref{actstd}) applied to the standard basis vectors of $[\lambda_1]$ and  $[\lambda_2]$ respectively, we have the action of the generators of $S_{n_1} \times S_{n_2}$ on the basis vectors $|\lambda_1, m_1 \rangle \otimes |\lambda_2, m_2 \rangle$.

%-------------------------------------------------------------------------------------------------------

\subsection{Subduction coefficients}
The \emph{subduction coefficients} (SDCs) are the entries of the matrix transforming between split and standard basis. Let $[\lambda_1] \otimes [\lambda_2]$ be a \emph{fixed} irrep of $S_{n_1}  \times S_{n_2}$ in $[\lambda] \downarrow S_{n_1} \times S_{n_2}$ and $| \lambda_1, \lambda_2 ; m_1, m_2 \rangle_{\eta}$  a generic vector of the split basis (where $m_1$ and $m_2$ are standard Young tableaux with Ferrers diagram $\lambda_1$ and $\lambda_2$ respectively,  and $\eta$ is the multiplicity label). We may expand such vectors in terms of the standard basis vectors $|\lambda ; m \rangle $ of $[\lambda]$:
\begin{equation}
| \lambda_1, \lambda_2 ; m_1, m_2 \rangle_{\eta} = \sum_m \ |\lambda ; m \rangle \langle \lambda ; m | \lambda_1, \lambda_2 ; m_1, m_2 \rangle_{\eta}. 
\end{equation}
Thus $\langle \lambda ; m | \lambda_1, \lambda_2 ; m_1, m_2 \rangle_{\eta}$ are the SDCs of $[\lambda] \downarrow [\lambda_1] \times [\lambda_2]$ with given multiplicity label $\eta$.
\par
Because the standard and the split basis vectors are orthogonal, the SDCs satisfy the following unitary conditions
\begin{equation}
\sum_m \ \langle \lambda ; m | \lambda_1, \lambda_2 ; m_1, m_2 \rangle_{\eta} \ \langle \lambda ; m | \lambda_1, \lambda'_2 ; m_1, m_2 \rangle_{\eta'} = \delta_{\lambda_2 \lambda'_2} \delta_{m_2 m'_2} \delta_{\eta \eta'}
\label{orton1}
\end{equation}
\begin{equation}
\sum_{\lambda_2 m_2 \eta} \ \langle \lambda ; m | \lambda_1, \lambda_2 ; m_1, m_2 \rangle_{\eta} \ \langle \lambda ; m' | \lambda_1, \lambda_2 ; m_1, m_2 \rangle_{\eta} = \delta_{m m'}.
\label{orton2}
\end{equation}
Notice that in (\ref{orton1}) we impose orthonormality between two different copies of multiplicity. It is not necessary, but it is the most natural choice. On the other hand, it imposes a precise and explicit form for the SDCs (see section $5$).

%-------------------------------------------------------------------------------------------------------  
\subsection{Subduction matrix and subduction space}
Using the linear equation method proposed by Chen and Pan~\cite{Chen} for Hecke algebras we may construct a matrix in such a way that the SDCs are the components of the kernel basis vectors.
\par
From (\ref{actsplit}), for $l \in \{1, 2, \ldots, n_1 - 1 \}$, we get
\begin{equation}
\langle \lambda ; m | g_l | \lambda_1, \lambda_2, m_1, m_2 \rangle = \langle \lambda ; m | (g_l | \lambda_1, m_1 \rangle) \otimes | \lambda_2, m_2 \rangle
\label{r1}
\end{equation}
and, writing $| \lambda_1, \lambda_2, m_1, m_2 \rangle_{\eta}$ and $g_l | \lambda_1, m_1 \rangle$ in the $S_n$-basis and $S_{n_1}$-basis respectively, (\ref{r1}) becomes
\begin{equation}
\sum_p \ \langle \lambda; m | g_l | \lambda; p \rangle \langle \lambda ; p | \lambda_1, \lambda_2 ; m_1, m_2 \rangle  = \sum_q \ \langle \lambda_1; q | g_l | \lambda_1; m_1  \rangle \langle \lambda ; m | \lambda_1, \lambda_2 ; q , m_2 \rangle.
\label{lem1}
\end{equation}
In an analogous way, for $l \in \{n_1 + 1, n_1 + 2, \ldots, n - 1 \}$, we get
\begin{equation}
\sum_p \ \langle \lambda; m | g_l | \lambda; p \rangle \langle \lambda ; p | \lambda_1, \lambda_2 ; m_1, m_2 \rangle  = \sum_q \ \langle \lambda_2; q | g_l | \lambda_2; m_2  \rangle \langle \lambda ; m | \lambda_1, \lambda_2 ; m_1 , q \rangle.
\label{lem2}
\end{equation}
\par 
Then, once we know the explicit action of the generators of $S_{n_1} \times S_{n_2}$ on the standard basis, (\ref{lem1}) and (\ref{lem2}) (written for $l \in \{1, \ldots, n_1-1, n_1+1, \ldots, n-1   \}$ and all standard Young tableaux $m$, $m_1$, $m_2$ with Ferrers diagrams $\lambda$, $\lambda_1$ and $\lambda_2$ respectively) define a linear equation system of the form:
\begin{equation}
\Omega (\lambda; \lambda_1, \lambda_2) \ \chi = 0  
\label{subdeq}
\end{equation}
where $\Omega (\lambda; \lambda_1, \lambda_2)$ is the \emph{subduction matrix} and $\chi$ is a vector with components given by the SDCs of $[\lambda] \downarrow [\lambda_1] \otimes [\lambda_2]$. We call the space of the solutions of (\ref{subdeq}), i.e. $\ker \ \Omega(\lambda; \lambda_1, \lambda_2)$, \emph{subduction space}.

%-------------------------------------------------------------------------------------------------------

\subsection{Explicit form for the subuction matrix}
Denoting as $N$, $N_1$ and $N_2$ the dimensions of the irreps $[\lambda]$, $[\lambda_1]$ and $[\lambda_2]$ respectively, (\ref{subdeq}) is a linear equation system with $N N_1 N_2$ unknowns (the SDCs) and $(n-2) N N_1 N_2$ equations. Thus $\Omega (\lambda; \lambda_1, \lambda_2)$ is a rectangular $(n-2) N N_1 N_2 \times N N_1 N_2$ matrix with real entries. 
Using the explicit action of $g_i$ given by (\ref{actstd}), we see that all equations of (\ref{subdeq}) have the form
\begin{equation}
\alpha^{(i)}_{m,m_{12}} \langle \lambda;  m | \lambda_1, \lambda_2 ; m_1, m_2 \rangle - \beta^{(i)}_m  \langle \lambda ;  g_i(m) | \lambda_1, \lambda_2 ; m_1, m_2 \rangle + 
\nonumber
\end{equation}
\begin{equation}
+ \beta^{(i)}_{m_{12}} \langle \lambda ;  m| \lambda_1 , \lambda_2 ; g_i(m_1),m_2 \rangle = 0 \ \ \ \ \ \ \  \text{if $i \in \{ 1, \ldots, n_1 - 1 \}$ },
\end{equation}
\begin{equation}
\alpha^{(i)}_{m,m_{12}} \langle \lambda;  m | \lambda_1, \lambda_2 ; m_1, m_2 \rangle - \beta^{(i)}_m  \langle \lambda ;  g_i(m) | \lambda_1, \lambda_2 ; m_1, m_2 \rangle + 
\nonumber
\end{equation}
\begin{equation}
+ \beta^{(i)}_{m_{12}} \langle \lambda ;  m| \lambda_1 , \lambda_2 ; m_1,g_i(m_2) \rangle = 0 \ \ \ \ \ \ \  \text{if $i \in \{ n_1 + 1, \ldots, n - 1 \}$ }
\label{sueq}
\end{equation}
where 
\begin{equation}
 \alpha^{(i)}_{m,m_{12}} = \frac{1}{d_i(m_{12})} - \frac{1}{d_i(m)} 
\end{equation}
\begin{equation}
\beta^{(i)}_m = \sqrt{1-\frac{1}{d_i^2(m)}}
\end{equation}
\begin{equation}
\beta^{(i)}_{m_{12}} = \sqrt{1-\frac{1}{d_i^2(m_{12})}}
\end{equation}
Notice that, by definition, 
\begin{equation}
d_i(m_{12}) =
\left\{
\begin{array}{cc}
d_i(m_1) & \text{if $i < n_1$} \\  
d_i(m_2) & \text{if $i > n_1$}
\end{array}
\right..
\end{equation}

%*******************************************************************************************************

\section{Subduction graph}

\begin{figure}
$
\xymatrix{
 1 & *{\bullet} \ar@{-}[rd]^(.39){(4)} & *{\bullet} \ar@{-}[ld] & *{\bullet} \ar@{-}[d]^{(4)}  \\    
 2 &   *{  \bullet} & *{\bullet} & *{\bullet}   \\
 3 & *{ \bullet} \ar@{-}[r]^{(4)} & *{\bullet} & *{\bullet} \\
 4 & *{ \bullet} \ar@{-}[r]^{(4)} & *{\bullet} & *{\bullet} \\
   & (1,1 ) & (1,2 ) & (1, 3)
}
$
\caption{\small{$4$-layer relative to the partitions $([4,1]; [1], [3,1])$. Nodes have coordinates given by the lexicografic ordering for Young tableaux with Ferrer diagram $[4,1]$ and for pairs of Young tableaux with Ferrer diagram $([1],[3,1])$. Two distinct $4$-coupled nodes are joined by an edge.}}
\label{ilayerex}
\end{figure}
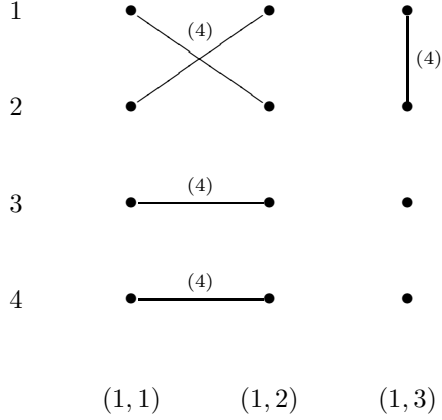

Given two standard Young tableaux $m_1$ and $m_2$ with the same Ferrers diagram, we say that they are \emph{$i$-coupled} if $m_1 = m_2$ or if $m_1=g_i(m_2)$. \par 
If $m_{12}=(m_1, m_2)$ is a pair of Standard Young tableaux with $k_1$ and $k_2$ boxes respectively, where $m_1$ is  filled by integers from $1$ to $k_1$ and $m_2$ from $k_1+1$ to $k_1 + k_2$, we define 
\begin{equation}
g_i(m_{12}) = \left\{
\begin{array}{cc}
(g_i(m_1),m_2) & \text{if $i < k_1$} \\
(m_1, g_i(m_2))& \text{if $i > k_1$}
\end{array}
\right.
\end{equation}
(notice that the action is not defined for $ i=k_1 $ because $g_{k_1}$ is not a generator of $S_{k_1} \times S_{k_2}$).  
Thus, denoting as $m_{34}$ another pair $(m_3,m_4)$, we say that $m_{12}$ and $m_{34}$ are \emph{$i$-coupled} if $m_{12} = m_{34}$ or if $g_i(m_{12}) = m_{34}$. \par
Let us now consider the three partitions $(\lambda; \lambda_1, \lambda_2)$ of $k$, $k_1$ and $k_2$ respectively, with $k_1 + k_2 = k$. We call \emph{node} each ordered sequence of three standard Young tableaux $(m; m_1, m_2)$ with Ferrers diagrams $\lambda$, $\lambda_1$ and $\lambda_2$ respectively and filled as described in the previous section. We denote it as $\langle m; m_{12} \rangle$. 
\par
The set of all nodes of $(\lambda; \lambda_1, \lambda_2)$ is called \emph{subduction grid} (or simply \emph{grid}). In analogy with the case of standard Young tableaux, we may define the action of $g_i$ on a node $n = \langle m; m_{12} \rangle $ as
\begin{equation}
g_i(n)= \langle g_i(m); g_i(m_{12}) \rangle.    
\end{equation}
Then we say that two nodes $n_1$ and $n_2$ are \emph{i-coupled} if $n_1 = n_2$ or if $n_1 = g_i(n_2)$. 
Once $i$ is fixed, it is easy to see that the $i$-coupling is an equivalence relation on the grid. Furthermore there are only four possible \emph{coupling configurations} between nodes:
\begin{enumerate}
\item one node $n=\langle m;m_{12} \rangle$ is called \emph{singlet} if $m=g_i(m)$ and if $m_{12}=g_i(m_{12})$;
\item two distinct $i$-coupled nodes $n=\langle m;m_{12} \rangle$ and  $n'=\langle m' ; m'_{12} \rangle$ are called \emph{vertical bridge} if $m_{12}=m'_{12}$;
\item two distinct $i$-coupled nodes $n=\langle m;m_{12} \rangle$ and  $n'=\langle m' ; m'_{12} \rangle$ are called \emph{horizontal bridge} if  $m=m'$;
\item four distinct nodes $n=\langle m;m_{12} \rangle$, $n'=\langle m';m'_{12} \rangle$, $n''=\langle m'';m''_{12} \rangle$ and $n'''= \langle m''';m'''_{12} \rangle$ such that $n=g_i(n')$ and $n''=g_i(n''')$ are called \emph{crossing} if $m \neq m'$, $m_{12} \neq m'_{12}$, $m'' \neq m''$ and $m''_{12} \neq m'''_{12}$.  
\end{enumerate}
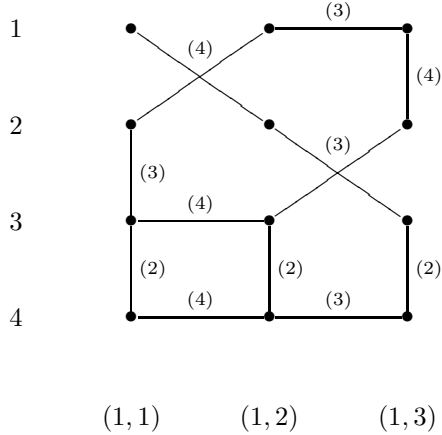
\begin{figure}
$\xymatrix{
 1 & *{\bullet} \ar@{-}[rd]^(0.39){(4)} & *{\bullet} \ar@{-}[ld] \ar@{-}[r]^{(3)} & *{\bullet}  \ar@{-}[d]^{(4)}  \\    
 2 &   *{  \bullet} \ar@{-}[d]^{(3)} & *{\bullet} \ar@{-}[rd]^(0.39){(3)} & *{\bullet} \ar@{-}[ld]  \\
 3 & *{ \bullet} \ar@{-}[r]^{(4)} \ar@{-}[d]^{(2)}  & *{\bullet} \ar@{-}[d]^{(2)} & *{\bullet} \ar@{-}[d]^{(2)} \\
 4 & *{ \bullet} \ar@{-}[r]^{(4)} & *{\bullet}\ar@{-}[r]^{(3)} & *{\bullet} \\
   & (1,1 ) & (1,2 ) & (1, 3)}$
\caption{\small{Subduction graph relative to $([4,1]; [1], [3,1])$. It is obtained by the overlap of the $2$-layer, $3$-layer and $4$-layer. Each $i$-layer can be distinguished by the label $(i)$ on the edges.}}
\label{subdgraf}
\end{figure}
\par 
The partition of the grid related to the $i$-coupling relation is called \emph{i-layer}. For each configuration it can be convenient to choose a representative node which we call \emph{pole}. Given a pole $p$ we denote by $\Gamma^{(i)}(p)$ the set of all nodes in its coupling configuration. For example, 
in figure~\ref{ilayerex} we show a graphic representation of the $4$-layer for $([4,1]; [1], [3,1])$. 
The nodes form a grid and their coordinates are obtained by the ordering number of the relative standard Young tableau (for example the lexicographic ordering~\cite{Chenbook}). Because each equivalence class is composed at most by two distinct nodes, we represent them as joined by an edge with a label for $i$. By convention, we choose the node on the top and left of the configuration as pole. We can see that $\{ \langle 1; 1, 1 \rangle, \langle 2; 1, 2 \rangle,  \langle 1; 1, 2 \rangle ,  \langle 2; 1, 1 \rangle \}$ is a crossing, $\{ \langle 1; 1, 3 \rangle, \langle 2; 1, 3 \rangle \}$ is a vertical bridge, $\{ \langle 3; 1, 1 \rangle, \langle 3; 1, 2 \rangle \}$ is an example of horizontal bridge and $\{ \langle 2; 1, 3 \rangle \}$ a singlet one.
\par
We call \emph{subduction graph} relative to $(\lambda; \lambda_1, \lambda_2)$ the \emph{overlap} of all $i$-slides (by overlap between two graphs we mean the graph obtained by identification of the corresponding  nodes). More simply, two distinct nodes $n$ and $n'$ of the grid are connected by an edge of the subduction graph if $n=g_i(n')$ for some $i$ (notice that if $n$ and $n'$ are $i$-coupled and $j$-coupled, then $i=j$). In figure~\ref{subdgraf} the subduction graph for $([4,1]; [1], [3,1])$ obtained from the overlap of the $2$-layer, the $3$-layer and the $4$-layer is shown.

%*******************************************************************************************************

\section{Solution space}
\subsection{Configurations and solutions}
The solution of (\ref{subdeq}) can be seen as an intersection of the $n-2$ subspaces of $\mathbb{R}^{N N_1 N_2}$ described by
\begin{equation}
\Omega^{(i)}(\lambda; \lambda_1, \lambda_2) \chi = 0,
\label{subdeqr}
\end{equation}
with $i \in \{1, \ldots, n_1 - 1, n_1+1, \ldots, n-1 \}$. We now construct an explicit solution of (\ref{subdeqr}), for a fixed $i$, by using the concept of $i$-layer.
\par
It is clear that we can associate each SDC of $[\lambda]\downarrow [\lambda_1] \otimes [\lambda_2]$ to a node of $(\lambda; \lambda_1, \lambda_2)$ in a one-to-one correspondence. 
Supposed $ p = \langle m; m_{12} \rangle$ as a fixed pole of a crossing configuration and $\Gamma^{(i)}(p)$ the set of all nodes of such a configuration, the solutions of the equations (\ref{subdeqr}), written for each $n \in \Gamma^{(i)}(p)$, are the kernel vectors of the matrix    \begin{equation} 
\Omega^{(i)}_{m;m_{12}} = 
\left(
\begin{array}{cccc}
\alpha^{(i)}_{m,m_{12}} & -\beta^{(i)}_m & \beta^{(i)}_{m_{12}} & 0 \\
-\beta^{(i)}_{g_i(m)} & \alpha^{(i)}_{g_i(m),m_{12}} & 0 & \beta^{(i)}_{m_{12}} \\
\beta^{(i)}_{g_i(m_{12})} & 0 & \alpha^{(i)}_{m,g_i(m_{12})} & -\beta^{(i)}_m \\
0 & \beta^{(i)}_{g_i(m_{12})} & -\beta^{(i)}_{g_i(m)} & \alpha^{(i)}_{g_i(m),g_i(m_{12})}
\end{array}
\right),
\label{subdeqrid}
\end{equation}
where the following relations hold:
\begin{equation}
\begin{array}{cc}
\alpha^{(i)}_{m,m_{12}} = - \alpha^{(i)}_{g_i(m),g_i(m_{12})}, \ & \alpha^{(i)}_{g_i(m),m_{12}} = - \alpha^{(i)}_{m,g_i(m_{12})}, \\
\beta^{(i)}_m = \beta^{(i)}_{g_i(m)}, \ & \beta^{(i)}_{m_{12}} = \beta^{(i)}_{g_i(m_{12})}
\end{array}
\label{sim}
\end{equation}
(they directly discend from $d_i(m) = - d_i(g_i(m))$ and $d_i(m_{12}) = - d_i(g_i(m_{12}))$). If we put 
\begin{equation}
\rho^{(i)}_m = \left( 
\begin{array}{cc}
\cos \theta^{(i)}_m & \sin \theta^{(i)}_m \\
\sin \theta^{(i)}_m & -\cos \theta^{(i)}_m
\end{array}
\right),  \ \ \ \ \   \cos \theta^{(i)}_m = \frac{1}{d_i(m)}, \ \  \sin \theta^{(i)}_m = \beta^{(i)}_m,
\label{ro1}
\end{equation}
\begin{equation}
\rho^{(i)}_{m_{12}} = \left( 
\begin{array}{cc}
\cos \theta^{(i)}_{m_{12}} & \sin \theta^{(i)}_{m_{12}} \\
\sin \theta^{(i)}_{m_{12}} & -\cos \theta^{(i)}_{m_{12}}
\end{array}
\right),  \ \ \ \ \   \cos \theta^{(i)}_{m_{12}} = \frac{1}{d_i(m_{12})}, \ \  \sin \theta^{(i)}_{m_{12}} = \beta^{(i)}_{m_{12}},
\label{ro2} 
\end{equation}
and we remember (\ref{sim}), then (\ref{subdeqrid}) can be written as 
\begin{equation}
\Omega^{(i)}_{m,m_{12}} =  \1 \otimes \rho^{(i)}_{m_{12}} - \rho^{(i)}_m \otimes \1 ,
\end{equation}
where $\1$ denotes the $2 \times 2$ identity matrix. It is straightforward that the kernel of $\Omega^{(i)}_{m,m_{12}}$ is generated by the vectors $e^{(i)}_m \otimes e^{(i)}_{m_{12}}$ and $\bar{e}^{(i)}_m \otimes \bar{e}^{(i)}_{m_{12}}$; here $e^{(i)}_m$ and $e^{(i)}_{m_{12}}$ are the eigenvectors of $\rho^{(i)}_m $ and $\rho^{(i)}_{m_{12}}$ respectively with eigenvalue $1$, while $\bar{e}^{(i)}_m$ and $\bar{e}^{(i)}_{m_{12}}$ are the corresponding ones with eigenvalue $-1$; from (\ref{ro1}) and (\ref{ro2}) we get
\begin{equation}
e^{(i)}_m =  \left(
\begin{array}{c}
\cos \frac{\theta^{(i)}_m}{2} \\ \sin \frac{\theta^{(i)}_m}{2}
\end{array} \right), \ \ \ \ \  e^{(i)}_{m_{12}} = \left(
\begin{array}{c}
\cos \frac{\theta^{(i)}_{m_{12}}}{2} \\ \sin{\frac{\theta^{(i)}_{m_{12}}}{2}}
\end{array} \right),
\end{equation}
and
\begin{equation}
\bar{e}^{(i)}_m = \left(
\begin{array}{c}
 - \sin \frac{\theta^{(i)}_m}{2} \\ \cos \frac{\theta^{(i)}_m}{2}
\end{array} \right) , \ \ \ \ \  \bar{e}^{(i)}_{m_{12}} =  \left(
\begin{array}{c}
- \sin \frac{\theta^{(i)}_{m_{12}}}{2} \\ \cos{\frac{\theta^{(i)}_{m_{12}}}{2}}
\end{array} \right)
\end{equation}
\par
In the case of vertical bridge configuration, we have $\beta^{(i)}_{m_{12}} = 0$ in (\ref{subdeqrid}). Therefore we can write
\begin{equation}
\Omega^{(i)}_{m,m_{12}} = (d_i(m_{12})\1 - \rho^{(i)}_m) \oplus (d_i(m_{12})\1 - \rho^{(i)}_m). 
\end{equation}
From $m_{12} = g_i(m_{12})$ it follows that we may only consider one of the two identical copies, thus
\begin{equation}
\Omega^{(i)}_{m,m_{12}} = d_i(m_{12})\1 - \rho^{(i)}_m .
\end{equation}
So, $\ker \Omega^{(i)}_{m,m_{12}}$ is generated by the eigenvector $e^{(i)}_m$ if $d_i(m_{12})=1$, by the eigenvector $\bar{e}^{(i)}_m$ if $d_i(m_{12})=-1$.
\par 
In an analogous way for a horizontal bridge we have $\beta^{(i)}_{m} = 0$ in (\ref{subdeqrid}). By the change of basis
\begin{equation}
\left( 
\begin{array}{cccc}
1 & 0 & 0 & 0 \\
0 & 0 & 1 & 0 \\
0 & 1 & 0 & 0 \\
0 & 0 & 0 & 1 
\end{array}
\right)  
\end{equation}
and using $m = g_i(m)$, we get
\begin{equation}
\Omega^{(i)}_{m,m_{12}} =  \rho^{(i)}_{m_{12}} - d_i(m)\1 .
\end{equation}  
Here $\ker \Omega^{(i)}_{m,m_{12}}$ is generated by the eigenvector $e^{(i)}_{m_{12}}$ if $d_i(m)=1$, by $\bar{e}^{(i)}_{m_{12}}$ if $d_i(m)=-1$.
\par
Finally, the case of singlet configuration is trivial because $\Omega^{(i)}_{m,m_{12}}$ is in diagonal form (both $\beta^{(i)}_{m}$ and $\beta^{(i)}_{m_{12}}$ are $0$). We can have two possibilities:
\begin{equation}
\Omega^{(i)}_{m,m_{12}} = ( 0 )
\end{equation}
or
\begin{equation}
\Omega^{(i)}_{m,m_{12}} = ( \pm 2 ).
\end{equation}
The kernel is the one-dimensional space generated by the vector $\{ 1\}$ or it is the trivial space.
\par 
All these results are summarized  in table 1, where with we deal with the various configurations, the coefficients of the linear subduction equations, their {\it $\Omega$ matrices} and the solution for the kernel vectors. Notice that, for the crossing configuration we distinguish the case $\alpha_{m;m_{12}} \neq 0$ from the case $\alpha_{m;m_{12}} = 0$. In the latter case we draw one of the edges with a dashed line. Furthermore, in the singlet configuration, we mark the trivial kernel solution by a label $0$ near the node. 
\begin{table}
\label{tabconf}
\begin{center}
\begin{tabular}{cccccc}
\toprule
Configuration & $\alpha_{m;m_{12}}$ & $\beta_m$ & $\beta_{m_{12}}$ & $\Omega_{m; m_{12}}$  & Basis  \\
\midrule

Crossing & & & & & \\
  $ 
\xymatrix{
  *{\bullet} \ar@{-}[rd] & *{\bullet} \ar@{-}[ld]   \\     
   *{  \bullet} & *{\bullet}    
 }
$   & $   
 \ne 0  $ & $\ne 0 $ &$\ne 0$ & $ 
\begin{array}{c}
\1 \otimes \rho_{m_{12}} \ + \\  - \ \rho_m \otimes \1
\end{array}
$ & 
$\begin{array}{c}
 e_m \otimes e_{m_{12}} \\
\bar{e}_m \otimes \bar{e}_{m_{12}} 
\end{array}$ \\
 $ 
\xymatrix{
  *{\bullet} \ar@{--}[rd] & *{\bullet} \ar@{-}[ld]   \\     
   *{  \bullet} & *{\bullet}    
 }
$ & $0$ & $
\begin{array}{c}
\beta \ne 0 \\  
\end{array}
$  &
 $
\begin{array}{c}
\beta \ne 0 \\  
\end{array}
$
  &$
\begin{array}{c}
\1 \otimes \rho \ +\\

 - \  \rho \otimes \1
\end{array}$
 & $\begin{array}{c}
 e \otimes e \\
\bar{e} \otimes \bar{e}
\end{array}$ 
\\ \midrule 

Vertical Bridge & & & & & \\
$ 
\xymatrix{
  *{\bullet} \ar@{-}[d]    \\     
   *{\bullet}     
 }$ & $\ne 0$ & $\ne 0$ & $0$ & $ \begin{array}{c} \1 -\rho_m \\ -\1 -\rho_m \end{array} $ & $\begin{array}{c} e_m  \\ \bar{e}_m \end{array}$ \\ 
\midrule
Horizontal Bridge & & & & & \\
 $ 
\xymatrix{
  *{\bullet} \ar@{-}[r] &        
   *{  \bullet} \\    
 }$   & $\ne 0 $ &$ 0$  &$\ne 0$ &$\begin{array}{c} \rho_{m_{12}} - \1 \\ \rho_{m_{12}} + \1 \end{array} $ & $ \begin{array}{c} e_{m_{12}} \\ \bar{e}_{m_{12}} \end{array}$ \\
\midrule 
Singlet & & & & & \\ 
$ 
\xymatrix{
 *{\bullet}     
 }$  & $0$ & $  0 $ & $ 0 $ & $(0 )$ & $1$ \\
 $ 
\xymatrix{
*{ \bullet_0}     
 }$   & $ \pm 2$ & $0$ & $0$ & $( \pm 2 )$ & - \\
\bottomrule 
\end{tabular}
\end{center}
\caption{\small{Fundamental $i$-coupling configurations, $\Omega$ matrices and solution space bases.}}
\end{table}

%-------------------------------------------------------------------------------------------------------

\subsection{Poles and their equivalence}
We will now prove that $\Omega^{(i)}_n$, with $n \in \Gamma^{(i)}(p)$, are equivalent up to change of basis that exchanges the nodes of the configuration. In this way, only the equations relative to one node of the configuration (the pole) are needed in the subduction system.
\par
Let us consider the crossing configuration. We first notice that 
\begin{equation}
\begin{array}{cc}
\rho^{(i)}_{g_i(m)} = \epsilon \rho^{(i)}_m \epsilon, &  \rho^{(i)}_{g_i(m_{12})} = \epsilon \rho^{(i)}_{m_{12}} \epsilon ,
\end{array}
\end{equation}
where 
$
\epsilon = \left ( \begin{array}{cc}
0 & 1 \\
1 & 0  
\end{array} \right)
$.
Then, observing that $\epsilon^2 = \1$, for the other three choices of pole we have
\begin{equation}
\begin{array}{c}
\Omega^{(i)}_{g_i(m),g_i(m_{12})} =  \1 \otimes \rho^{(i)}_{g_i(m_{12})} - \rho^{(i)}_{g_i(m)} \otimes \1 = \\
= \1 \otimes \epsilon \rho^{(i)}_{m_{12}} \epsilon - \epsilon \rho^{(i)}_m \epsilon \otimes \1 = (\epsilon \otimes \epsilon) (\1 \otimes \rho^{(i)}_{m_{12}} - \rho^{(i)}_{m} \otimes \1) (\epsilon \otimes \epsilon) = \\
= (\epsilon \otimes \epsilon) \Omega^{(i)}_{m,m_{12}} (\epsilon \otimes \epsilon);
\end{array}
\end{equation}
\begin{equation}
\begin{array}{c}
\Omega^{(i)}_{m,g_i(m_{12})} =  \1 \otimes \rho^{(i)}_{g_i(m_{12})} - \rho^{(i)}_{m} \otimes \1 = \\
= \1 \otimes \epsilon \rho^{(i)}_{m_{12}} \epsilon -  \rho^{(i)}_m  \otimes \1 = (\1 \otimes \epsilon) (\1 \otimes \rho^{(i)}_{m_{12}} - \rho^{(i)}_{m} \otimes \1) (\1 \otimes \epsilon) = \\
= (\1 \otimes \epsilon) \Omega^{(i)}_{m,m_{12}} (\1 \otimes \epsilon);
\end{array}
\end{equation}
\begin{equation}
\begin{array}{c}
\Omega^{(i)}_{g_i(m),m_{12}} =  \1 \otimes \rho^{(i)}_{m_{12}} - \rho^{(i)}_{g_i(m)} \otimes \1 = \\
= \1 \otimes \rho^{(i)}_{m_{12}}  - \epsilon \rho^{(i)}_m \epsilon \otimes \1 = (\epsilon \otimes \1) (\1 \otimes \rho^{(i)}_{m_{12}} - \rho^{(i)}_{m} \otimes \1) (\epsilon \otimes \1) = \\
= (\epsilon \otimes \1) \Omega^{(i)}_{m,m_{12}} (\epsilon \otimes \1).
\end{array}
\end{equation}
In any case we are able to find the suitable change of basis.
\par
Of course, for the bridge configurations the change of pole is equivalent to a change of basis by $\epsilon$. The singlet configuration is a trivial case. 

%-------------------------------------------------------------------------------------------------------

\subsection{Structure of the subduction space}
We can now write the explicit solution space $\chi^{(i)}$ for (\ref{subdeqr}) as a suitable subspace of $\mathbb{R}^N \otimes \mathbb{R}^{N_1 N_2}$.  
If we define the vectors (in components)
\begin{equation}
(\lambda^{(i)}_m)_k = \left\{
\begin{array}{cc}
 0 & \text{if $k$ is not $i$-coupled with $m$}  \\
 (e^{(i)}_m)_k & \text{if $k$ is $i$-coupled with $m$}
\end{array} \right.
\end{equation}
\begin{equation}
(\bar{\lambda}^{(i)}_m)_k = \left\{
\begin{array}{cc}
 0 & \text{if $k$ is not $i$-coupled with $m$}  \\
 (\bar{e}^{(i)}_m)_k & \text{if $k$ is $i$-coupled with $m$}
\end{array} \right.
\end{equation}
\begin{equation}
(\delta_m)_k = \left\{
\begin{array}{cc}
 0 & \text{if $k \neq m$}  \\
 1 & \text{if $k = m$}
\end{array} \right.
\end{equation}
and the spaces
\begin{equation}
\chi^{(i)}_{m;m_{12}} = \left\{
\begin{array}{cc}
\langle \alpha^{(i)}_{m;m_{12}} \delta_m \otimes \delta_{m_{12}} \rangle & \text{if $d_i(m) = \pm 1$ and $d_i(m_{12}) = \pm 1$ } \\
\langle \lambda^{(i)}_{m} \otimes \delta_{m_{12}} \rangle & \text{if $d_i(m)\neq \pm 1$ and $d_i(m_{12})=  1$}  \\
\langle \bar{\lambda}^{(i)}_{m} \otimes \delta_{m_{12}} \rangle & \text{if $d_i(m)\neq \pm 1$ and $d_i(m_{12})=  -1$}  \\
\langle \delta_{m} \otimes \lambda^{(i)}_{m_{12}} \rangle & \text{if $d_i(m) =  1$ and $d_i(m_{12}) \neq \pm 1$}  \\
\langle \delta_{m} \otimes \bar{\lambda}^{(i)}_{m_{12}} \rangle & \text{if $d_i(m) =  -1$ and $d_i(m_{12}) \neq \pm 1$}  \\
\langle \lambda^{(i)}_{m} \otimes \lambda^{(i)}_{m_{12}} ,  \bar{\lambda}^{(i)}_{m} \otimes \bar{\lambda}^{(i)}_{m_{12}}   \rangle &  \text{if $d_i(m) \neq \pm 1$ and $d_i(m_{12}) \neq \pm 1$} 
\end{array}\right. ,
\label{basis}
\end{equation}
denoted by $P^{(i)}$ the set of the poles for the $i$-layer and observing that the set of the configurations for the $i$-layer is a partition of the grid, we have 
\begin{equation}
\chi^{(i)} = \bigoplus_{  (m; m_{12}) \in P^{(i)}}  \chi^{(i)}_{m; m_{12}}. 
\label{ilayerdec}
\end{equation}     
So the general solution of (\ref{subdeq}) is the intersection of $n-2$ subspaces, i.e. 
\begin{equation}
\chi = \bigcap_{i \in I} \chi^{(i)},  
\label{ilayerdec1}
\end{equation}
with $I= \{1, \ldots, n_1 -1, n_1 + 1, \ldots , n -1 \}$.
\par
Now we can outline an algorithm (in pseudo-code) to determine the SDCs for $[\lambda] \downarrow [\lambda_1] \otimes [\lambda_2]$: 
\begin{enumerate}
\item for $i \in I$ :
\begin{enumerate}
\item construct the $i$-layer;
\item choose poles;
\item for each pole (configuration):\\
%\begin{enumerate}
%\item 
construct the space $\chi^{(i)}_p$ by (\ref{basis});
%\end{enumerate} 
\item construct $ \chi^{(i)}$ by (\ref{ilayerdec});
\end{enumerate}
\item determine $\chi$ as intersection of all $ \chi^{(i)}$.  
\end{enumerate}
\par 
Step (ii) can be performed by using the subduction graph to obtain a minimal number of equations. In fact, one may associate a suitable equation deriving from (\ref{ilayerdec1}) to each edge (two for the crossing) of the graph (nodes represents the unknown SDCs). Then, starting from a suitable node in the graph, we can extract such equations by applying a graph searching algorithm which is able to reach every edge~\cite{Gibbons}. As regard it is useful to notice that equations associated to closed loops of bridge configurations are always linearly dependent.       

%*******************************************************************************************************

\section{Orthonormalization and form}
The subduction space given by (\ref{ilayerdec1}) has dimension $\mu$ equal to the multiplicity of $[\lambda] \downarrow [\lambda_1] \otimes [\lambda_2]$. Then SDCs are not univocally determined. A choice of orthonormality between the different copies of multiplicity imposes a precise form for the multiplicity separations. 
\par
Let $\{ \chi_1, \ldots, \chi_{\mu} \}$ be a basis in the subduction space. Orthonormality implies for the scalar products:
\begin{equation}
(\chi_{\eta} , \chi_{\eta'}) = N_1 N_2 \ \delta_{\eta \eta'}.
\label{on} 
\end{equation}
If we denote by $\chi$ the matrix which has the basis vectors of the subduction space as columns, we may orthonormalize it by a suitable $\mu \times \mu $ matrix $\sigma$, i.e. 
\begin{equation}
\tilde{\chi} = \chi \sigma.
\label{orton}   
\end{equation}
In (\ref{orton}) $\tilde{\chi}$ is the matrix which has the orthonormalized basis vectors of the subduction space as columns. Now we can write (\ref{on}) as
\begin{equation}
\sigma^t \ \tau \ \sigma = \1, 
\label{silv}
\end{equation}
where $\1$ is the $\mu \times \mu$ identity matrix and $\tau$ is the $\mu \times \mu$ positive defined quadratic form given by
\begin{equation}
\tau = \frac{1}{N_1 N_2} \ \chi^{t} \chi.
\end{equation}
From (\ref{silv}) we can see $\sigma$ as the Sylvester matrix of $\tau$, i.e. the matrix for the change of basis that reduces $\tau$ in the identity form. We can express $\sigma$ in terms of the orthonormal matrix $O_{\tau}$ that diagonalizes $\tau$
\begin{equation}
\sigma = O_{\tau} D^{-\frac{1}{2}}_\tau O,
\end{equation}
where $D^{-\frac{1}{2}}_{\tau}$ is the diagonal matrix with eigenvalues given by the inverse square root of the eigenvalues of $\tau$ and $O$ a { \it generic} orthogonal matrix. Thus, the general form for the orthonormalized $\chi$ is 
\begin{equation}
\tilde{\chi} = \chi O_{\tau} D^{-\frac{1}{2}}_\tau O. 
\label{genchi}
\end{equation}
\par
(\ref{genchi}) suggests some considerations on the form of the SDCs. First we notice that in case of multiplicity-free subduction, only one choice of global phase has to be made (for example Young-Yamanouchi phase convention~\cite{Chenbook}). It derives from the trivial form of the orthogonal $1 \times 1$ matrices $O$ and $O_{\tau}$.
\par
In the general case of multiplicity $\mu > 1$, $2^{\mu - 1}$ phases deriving from the $O_{\tau}$ matrix and $1$ phase from the matrix $O$ have to be fixed. Therefore we have $2^{\mu - 1} + 1$ phases to choose. Furthermore we have other $\frac{\mu (\mu -1)}{ 2}$ degrees of freedom deriving from $O$. In sum we have a total of $(2^{\mu - 1} + 1) + \frac{\mu (\mu -1)}{ 2} $ choices to make. We agree with~\cite{McAven} for the case of multiplicity $2$, in which we need three phases and one extra parameter to govern the multiplicity separation.
\par 
Other aspects have to be considered if we want to find the simplest and most natural form for these symmetric group transformation coefficients. In~\cite{McAven} the authors expose the following suitable requirements:
\begin{enumerate}
\item the trasformation coefficients should be chosen to be real if possible;
\item phases and the multiplicity separation should be chosen to be indipendent from $n$;
\item the multiplicity separation is to be chosen so that a maximal number of zero coefficients is obtained;
\item it is desirable to have the coefficients written as a single surd of the form $a \sqrt{b}/c$, with $a$, $b$, $c$ integers;
\item the prime numbers which occur in the surds should be as small as possible.   
\end{enumerate}
The first two statements are automatically verified if we assume (\ref{genchi}). The last three heavily depend on the form of $\tau$. This can be an interesting mathematical point to study (but its relevance is relative from a purely physical point of view). We think the form of eigenvalues and eigenvectors of $\tau$ are the only important factors in this regard. Non-normalized SDCs deriving from (\ref{ilayerdec1}) seem always to be in a {\it simple} form. 

%*******************************************************************************************************

\section{Conclusion}
In this paper we have investigated the linear equation method for symmetric groups, proposed by Chen {\it et al} for the determination of the SDCs as solution of a linear system. We have proven that such a system, which  is constituted by a complicated primal structure of dependent linear equations, can be simplified by choosing a minimal set of sufficient equations related to the concept of subduction graph. Furthermore, the subduction graph provides a very practical way to choose such equations and it suggests that subduction coefficients may be seen as a subspace of $\mathbb{R}^{N} \otimes \mathbb{R}^{N_1 N_2}$ obtained by the intersection of only $n-2$ explicit subspaces (each one in corrispondence with an $i$-layer) instead of the original $(n-2) N N_1 N_2$ ones. Consequently we have a more explicit insight into the structure of the standard to split basis transformation. 
\par 
We have proposed a general form for the SDCs resulting from the only requirement of orthonormality and  we have seen that the multiplicity separation can be described in terms of the Sylvester matrix of the positive defined quadratic form $\tau$ describing the scalar product in the subduction space. Then we are able to link the freedom in fixing the multiplicity separation to the freedom deriving from the choice of the Sylvester matrix. The number of phases and free factors for the general multiplicity separation can be expressed as function of the multiplicity $\mu$ (i.e. the dimension of the subduction space). It seems to be a crucial question if one may fix the Sylvester matrix to obtain all the requirements of {\it simplicity} given in the previous section for the form of each coefficient. We conjecture that such a form only depends on the form of the eigenvalues and eigenvectors of $\tau$.
\par 
We are going to implement a Mathematica code which uses the results in this paper to easily provide the SDCs relative to high dimension irreps. An interesting example is $[4,3,2,1] \downarrow [3,2,1] \otimes [3,1]$, because it represents the first case of symmetric groups subduction with multiplicity three and the corresponding SDCs are still unknown. Other aspects that could be investigated with interest are the following. First, the possibility of giving an explicit description of the intersection subspace (\ref{ilayerdec1}) to achieve a comprehensive algebraic solution of the subduction system. Second, the way to choose the Sylvester matrix to fix the multiplicity separation. Third, we think that the subduction graph approach can be useful to other subduction problems such as those related to Brauer algebras and quantum groups, which are important in many physical models. Moreover the results of this paper can be directly applied to the subduction problem in Hecke algebras~\cite{Chen}. 

%*******************************************************************************************************

\ack
The author is grateful to Massimo Campostrini for his valuable support. It is a pleasure to thank Massimo Mongia and Francesco Veneziano for very stimulating discussions and Martina Johnson, Giuseppe Della Sala and Davide Vittone for critical reading of this manuscript and useful suggestions.  

%*******************************************************************************************************


\section*{References}

\begin{thebibliography}{99}
\bibitem{Elliot} Elliot J P, Hope J and Jahn H A 1953 {\it Phil. Trans. R. Soc.} A $\mathbf{246}$ 241
%%CITATION=NONE;%%

\bibitem{Kramer} Kramer P 1968 {\it Z. Phys.} $\mathbf{216}$ 68-93
%%CITATION=NONE;%%
\bibitem{Vanagas} Vanagas V V 1971 {\it Algebraic Methods in Nuclear Theory} (Vilnius: Mintis) (in Russian)
%%CITATION=NONE;%%
\bibitem{Haase} Haase R W and Butler P H 1984 {\it J. Phys. A: Math. Gen} $\mathbf{17}$ 47-59
%%CITATION=JPAGB,17,47;%%

\bibitem{Chenbook} Chen J Q 1989 { \it Group Representation Theory for Physicists} (World  Scientific Publishing)
%%CITATION=NONE;%%
\bibitem{Pan1} Pan F 1993 { \it J. Phys. A: Math. Gen.} $\mathbf{26}$ 4621-4632
%%CITATION=JPAGB,26,4621;%% 

\bibitem{Kaplan2} Kaplan I G 1975 {\it Symmetry of Many-Electron Systems} (New York: Academic)
%%CITATION=NONE;%%
\bibitem{Rao} Suryanarayana C and Rao M K 1982 {\it J. Phys. A: Math. Gen} $\mathbf{15}$ 2013-2016
%%CITATION=JPAGB,15,2013;%%
\bibitem{McAven}  McAven L F, Butler P H and Hamel A M 1998 {\it J. Phys. A: Math. Gen.} $\mathbf{31}$ 8363-8372
%%CITATION=JPAGB,31,8363;%%

\bibitem {Horie} Horie H 1964 {\it J. Phys. Soc. Japan} $\mathbf{19}$ 1783-1799
%%CITATION=JUPSA,19,1783;%%
\bibitem{Kaplan1} Kaplan I G 1961 {\it Zh. Eksp. Teor. Fiz.} $\mathbf{41}$ 560 (Engl. transl. 1962 {\it Sov. Phys.-JETP} $\mathbf{14}$ 401-407)
%%CITATION=NONE;%%
\bibitem{Chen1} Chen J Q, Collinson D F and Gao M J 1983 {\it J. Math. Phys.} $\mathbf{24}$ 2695-2705
%%CITATION=NONE;%%1

\bibitem{Butler} Butler P H 1981 {\it Point Group Symmetry Applications: Method and Tables} (New York: Plenum)
%%CITATION=NONE;%%

\bibitem{McAven1} McAven L F and Butler P H 1999 {\it J. Phys. A: Math. Gen.} $\mathbf{32}$ 7509-7522
%%CITATION=JPAGB,32,7509;%%
\bibitem{McAven2} McAven L F and Hamel A M 2002 {\it J. Phys. A: Math. Gen. } $\mathbf{35}$ 1719-1725
%%CITATION=JPAGB,35,1719;%%

\bibitem{Chen} Pan F and Chen J Q 1993 { \it J. Phys. A: Math. Gen.} $\mathbf{26}$ 4299-4310
%%CITATION=JPAGB,26,4299;%%

\bibitem{Fulton} Fulton W 1997 { \it Young Tableaux With Applications to Representation Theory and Geometry} (Cambridge University Press - London Mathematical Society Student Texts $\mathbf{35}$)
%%CITATION=NONE;%%

\bibitem{Gibbons} Gibbons A 1994 {\it Algorithmic Graph Theory} (Cambridge University Press)
 %%CITATION=NONE;%%
\end{thebibliography}
\end{document}